\documentclass[10pt]{amsart}

\usepackage{amssymb,amsthm,amsmath}
\usepackage[numbers,sort&compress]{natbib}
\usepackage{color}
\usepackage{graphicx}
\usepackage{tikz}

\title[ ]{Sharp  bound on the largest  positive  eigenvalue   for one-dimensional Schr\"odinger operators}

\author{Wencai Liu}
\address[Wencai Liu]{Department of Mathematics, University of California, Irvine, California 92697-3875, USA} \email{liuwencai1226@gmail.com}

\theoremstyle{plain}
\newtheorem{theorem}{Theorem}[section]

\newtheorem{lemma}[theorem]{Lemma}

\theoremstyle{definition}

\newtheorem{remark}[theorem]{Remark}

\begin{document}


\begin{abstract}
Let $H=-D^2+V$  be a Schr\"odinger operator on  $ L^2(\mathbb{R})$, or on   $ L^2(0,\infty)$.
Suppose the potential satisfies  $\limsup_{x\to \infty}|xV(x)|=a<\infty$. We prove that $H$ admits no eigenvalue larger than
$ \frac{4a^2}{\pi^2}$. For any positive $a$ and $\lambda$  with $0<\lambda< \frac{4a^2}{\pi^2}$, we   construct   potentials  $V$
such that $\limsup_{x\to \infty}|xV(x)|=a $  and the associated Sch\"rodinger operator  $H=-D^2+V$  has eigenvalue $\lambda$.
\end{abstract}
\maketitle

 \section{Introduction and main results}

 Let $H=-D^2+V$ be a   one-dimensional
 Sch\"rodinger operator   on   $ L^2(\mathbb{R})$, or  on $ L^2(0,\infty)$, where the  potential $V$ satisfies
 \begin{equation*}
   \limsup_{x}|xV(x)|=a<\infty.
 \end{equation*}
 By the  classical Weyl theorem, the essential spectrum $\sigma_{\rm ess}(H)=[0,\infty)$.
In this paper, we  are interested in quantitative relation of    the largest positive eigenvalue   and $a$.
It is well known that there is no eigenvalue larger than $a^2$,  see paper \cite{kato} for example.
This implies there is no eigenvalue embedded into the essential spectrum if $V(x)=o(x^{-1})$ as $x$ goes to infinity.
As a standard example with  Wigner-von Neumann type potentials shows, for any $0<\lambda<\frac{a^2}{4}$,  there exists $V$ with $\limsup_{x}|xV(x)|=a$ such that $\lambda$ is an eigenvalue.
Thus $V(x)=o(x^{-1})$ is the spectral transition for existence of  eigenvalue embedded into the essential spectrum.
Naboko \cite{naboko} and Simon \cite{simon} constructed examples to show that dense point spectra can be embedded into essential spectrum if
the potential decreases  slightly slower than $V(x)=O(x^{-1})$.
The fact that  $V(x)=\frac{O(1)}{1+x}$ is the  spectral transition for singular continuous  embedding into the essential spectrum was established by Kiselev \cite{ki1}.

Kiselev-Last-Simon \cite{ki2} proved that
if $\limsup_{x}|xV(x)|=a$, then the sum of   eigenvalues $\lambda_n$ of $-D^2+V$ is finite, that is
   $ \sum \lambda_n\leq \frac{a^2}{2}$.

In this paper, we focus on  the largest eigenvalue and obtain the following sharp result.
 \begin{theorem}\label{thm1}
Suppose potential $V$  satisfies
\begin{equation*}
   \limsup_{x\to \infty}|xV(x)|=a<\infty.
 \end{equation*}
 Then $H$ admits no eigenvalue larger than
$ \frac{4 a^2}{\pi^2}$.
 \end{theorem}
 \begin{theorem}\label{thm2}
 For any positive  $a$ and $\lambda$  with $0<\lambda< \frac{4a^2}{\pi^2}$,  there exist      potentials  $V$
such that $\limsup_{x\to \infty}|xV(x)|=a $  and the  associated  Schr\"odinger operator  $H=-D^2+V$  has eigenvalue $\lambda$.
 \end{theorem}
 Our proof is based on the  modified Pr\"ufer transformation and some basic analysis. See \cite{ki2} for  more details about modified Pr\"ufer transformation.
 \begin{remark}\footnote{I would like to thank Barry Simon for telling me the full history of the problem.}
 After we finished this paper, we noticed that Theorem 1.2 has been proved  by Halvorsen  \cite{Hal}. See \cite{atk,eas,ani} for  more generalization.
 Later, Remling  also   addressed the problem  \cite{remling1998absolutely} and showed that  $H$ has purely absolute continuous spectrum in $ (\frac{4 a^2}{\pi^2},\infty)$.
 We refer readers to Simon's article for full history \cite{simon2017tosio}.
 \end{remark}
 \section{Proof of Theorems \ref{thm1} and \ref{thm2} }
 Without loss of generality, we only consider the half line case $L^2(0,\infty)$.

 Let $\lambda=k^2$ with $k>0$ and
 suppose $ u $ is a solution of
 \begin{equation}\label{Gei}
    -u^{\prime\prime}(x) +V(x)u(x)= k^2 u(x).
 \end{equation}
 Change variables to
 \begin{eqnarray*}
   u^{\prime}(x) &=&  kR(x)\cos(\theta(x)) \\
    u(x) &=&  R(x)\sin(\theta(x)).
 \end{eqnarray*}
 We get a pair of equations
 \begin{eqnarray}
   \frac{d\theta}{dx} &=& k-\frac{V(x)}{k}\sin^2\theta \label{Gtheta}\\
  \frac{d\log R}{dx}&=&  \frac{1}{2k} V(x)\sin2\theta.\label{GR}
 \end{eqnarray}
 If $V=0$, then $\theta(x)=\theta_0+kx$ and $R(x)=R_0$ is a solution.
 The following lemma is well known. The proof is basic, the readers can see Lemma 4.2 in  \cite{ki2} for the details.
 \begin{lemma}[Lemma 4.2, \cite{ki2}]\label{Lem1}
 If $u\in L^2(0,\infty)$ is a solution of \eqref{Gei}, then
 \begin{equation*}
    R(\cdot)\in L^2(0,\infty).
 \end{equation*}

 \end{lemma}
 {\textbf{Proof of Theorem \ref{thm1}}}

 \begin{proof}
 Under the assumption of Theorem \ref{thm1}, for any $\epsilon>0$,
 there exists $x_0>0$ such that  for all $x>x_0$
 \begin{equation*}
   | V(x)|\leq \frac{a+\epsilon}{1+x}.
 \end{equation*}
 In the following, we always assume $\epsilon>0$ is sufficiently small and may change even in the same formula.
 We also assume $x>0$ is large enough.

 By \eqref{GR}, one has
 \begin{equation}\label{e4}
   \log R(x)\geq \log R(x_0)- \frac{a+\epsilon}{2k}\int_{x_0}^x\frac{|\sin2\theta(y)|}{y+1}dy.
 \end{equation}

 Now we will estimate $\int_{x_0}^x\frac{|\sin2\theta(y)|}{y+1}dy$.
 Let $i_0$ be the largest positive integer such that $2\pi i_0<\theta(x_0)$.
 By \eqref{Gtheta},  there exist
  $x_0<x_1<x_2<\cdots<x_n<x<x_{n+1}$   such that
 \begin{equation*}
    \theta(x_i)= 2\pi i_0+i\frac{\pi}{2}
 \end{equation*}
 for $i=1,2,\cdots,n,n+1$.

 By \eqref{Gtheta}, one has
 \begin{equation*}
    |x_{i+1}-x_{i}|=\frac{\pi}{2k}+ \frac{O(1)}{x_i+1}.
 \end{equation*}
 Then, one has
 \begin{equation}\label{e2}
    n\leq (\frac{2k}{\pi}+\epsilon)x
 \end{equation}
 and
 \begin{equation}\label{e3}
    (\frac{\pi}{2k}-\epsilon)i<x_i-x_0<(\frac{\pi}{2k}+\epsilon)i.
 \end{equation}
 For $y\in[x_i,x_{i+1})$, we have
 \begin{eqnarray*}
    \theta(y) &=&2\pi i_0+i\frac{\pi}{2}+k(y-x_i)+\frac{O(1)}{1+x_i} \\
     &=& 2\pi i_0+i\frac{\pi}{2}+k(y-x_i)+\frac{O(1)}{1+i}.
 \end{eqnarray*}
 Which implies
 \begin{eqnarray}
   \nonumber \int_{x_i}^{x_{i+1}}|\sin2\theta(y)|dy &=& \int_{0}^{\frac{\pi}{2k}}\sin( 2ky )dy+\frac{O(1)}{1+i} \\
     &=&\frac{1}{k}+\frac{O(1)}{1+i}. \label{e1}
 \end{eqnarray}

By \eqref{e2}, \eqref{e3} and \eqref{e1}, we obtain
 \begin{eqnarray*}
  \int_{x_0}^x\frac{|\sin2\theta(y)|}{y+1}dy&\leq& \sum_{i=1}^{n+1}(\frac{1}{k}+\frac{O(1)}{1+i})\frac{2k}{(\pi-\epsilon)i}+O(1)\\
    &\leq&( \frac{2}{\pi}+\epsilon)\ln x+O(1).
 \end{eqnarray*}
Combining with \eqref{e4}, we have
 \begin{equation*}
    \log R(x)\geq -(\frac{a}{k\pi}+\epsilon)\ln x +O(1).
 \end{equation*}
 Assume $k>\frac{2a}{\pi}$, one has
 \begin{equation*}
    \frac{a}{k\pi}+\epsilon<\frac{1}{2}.
 \end{equation*}
 Thus
 \begin{equation*}
    R(x)\geq \frac{1}{\sqrt{x}}
 \end{equation*}
 for large $x$.  By Lemma \ref{Lem1}, $u\notin L^2 (0,\infty)$. Thus $\lambda=k^2$ is not an eigenvalue.

 \end{proof}
 {\textbf{Proof of Theorem \ref{thm2} }}
 \begin{proof}
 We define
 \begin{equation}\label{Def}
    V(x)=-\frac{a}{1+x}{\rm sgn}(\sin2\theta(x)),
 \end{equation}
 where ${\rm sgn}(\cdot) $ is the sign function.
 Substitute  \eqref{Def} into \eqref{Gtheta}, and
 solve the   nonlinear system for $\theta$ with the initial condition $\theta(0)=\theta_0$.
It is not difficult to  see that \eqref{Gtheta}   has a unique piecewise
smooth global solution by a standard ODE existence and uniqueness theorem.
Thus $V(x)$ is well defined and
\begin{equation}\label{e6}
    \frac{d\log R}{dx}= - \frac{a}{2k} \frac{\sin2\theta}{1+x}.
\end{equation}
Under the notations in the proof of Theorem \ref{thm1},
we have
 \begin{eqnarray*}
  \int_{1}^x\frac{|\sin2\theta(y)|}{y+1}dy&\geq& \sum_{i=1}^{n+1}(\frac{1}{k}+\frac{O(1)}{1+i})\frac{2k}{(\pi+\epsilon)i}+O(1)\\
    &\geq&( \frac{2}{\pi}-\epsilon)\ln x+O(1).
 \end{eqnarray*}
Combining with \eqref{e6},  it is easy to see
 \begin{equation*}
    \log R(x)\leq -(\frac{a}{k\pi}-\epsilon)\ln x +O(1).
 \end{equation*}
 Assume $k<\frac{2a}{\pi}$, then
 \begin{equation*}
    \frac{a}{k\pi}-\epsilon>\frac{1}{2}+\epsilon.
 \end{equation*}
 Thus
 \begin{equation*}
    R(x)^2\leq \frac{1}{x ^{1+\epsilon}},
 \end{equation*}
 for large $x$. This yields  $u\in L^2(0,\infty)$,  so $\lambda=k^2$ is an eigenvalue.

 \end{proof}
 \section*{Acknowledgments}

 The paper is inspired by Fan Yang's talk at  ergodic Schr\"odinger operators weekly graduate seminar at UCI, where she presented paper \cite{ki2}. I am appreciated her fascinating presentation.
 I would like to thank Svetlana Jitomirskaya for comments on earlier versions of the manuscript. I also
 thank Barry Simon for telling me  the full history of the problem (see footnote 1).
  The author  was supported by the AMS-Simons Travel Grant 2016-2018 and NSF DMS-1700314. This research was also
partially supported by NSF DMS-1401204.


\footnotesize
\begin{thebibliography}{10}

\bibitem{ani}
L.~Anikeeva.
\newblock The absence of solutions in ${L}^2 [x_0,\infty]$ of a second order
  linear differential equation.
\newblock {\em Differential Equations}, 6:1721--1724, 1970.

\bibitem{atk}
F.~Atkinson and W.~Everitt.
\newblock Bounds for the point spectrum for a {S}turm-{L}iouville equation.
\newblock {\em Proceedings of the Royal Society of Edinburgh Section A:
  Mathematics}, 80(1-2):57--66, 1978.

\bibitem{eas}
M.~S.~P. Eastham and H.~Kalf.
\newblock {\em Schr{\"o}dinger-type operators with continuous spectra},
  volume~65.
\newblock Pitman Publishing, 1982.

\bibitem{Hal}
S.~Halvorsen.
\newblock Sharp bounds and ${L}_p-$stability for solutions of secondorder
  linear ordinary differential equations.
\newblock {\em Report No. 9/75, Matematisk Institutt, Trondheim}, 1975.

\bibitem{kato}
T.~Kato.
\newblock Growth properties of solutions of the reduced wave equation with a
  variable coefficient.
\newblock {\em Communications on Pure and Applied Mathematics}, 12(3):403--425,
  1959.

\bibitem{ki1}
A.~Kiselev.
\newblock Imbedded singular continuous spectrum for {S}chr{\"o}dinger
  operators.
\newblock {\em Journal of the American Mathematical Society}, 18(3):571--603,
  2005.

\bibitem{ki2}
A.~Kiselev, Y.~Last, and B.~Simon.
\newblock Modified {P}r{\"u}fer and {E}{F}{G}{P} transforms and the spectral
  analysis of one-dimensional {S}chr{\"o}dinger operators.
\newblock {\em Communications in mathematical physics}, 194(1):1--45, 1998.

\bibitem{naboko}
S.~N. Naboko.
\newblock Dense point spectra of {S}chr{\"o}dinger and {D}irac operators.
\newblock {\em Theoretical and Mathematical Physics}, 68(1):646--653, 1986.

\bibitem{remling1998absolutely}
C.~Remling.
\newblock The absolutely continuous spectrum of one-dimensional
  {S}chr{\"o}dinger operators with decaying potentials.
\newblock {\em Communications in mathematical physics}, 193(1):151--170, 1998.

\bibitem{simon}
B.~Simon.
\newblock Some {S}chr{\"o}dinger operators with dense point spectrum.
\newblock {\em Proceedings of the American Mathematical Society},
  125(1):203--208, 1997.

\bibitem{simon2017tosio}
B.~Simon.
\newblock Tosio {K}ato's {W}ork on {N}on--{R}elativistic {Q}uantum {M}echanics.
\newblock {\em arXiv preprint arXiv:1711.00528}, 2017.

\end{thebibliography}
\end{document}